\documentstyle[preprint,aps]{revtex}

\begin{document}
\draft
\title{Null Result for the Violation of Equivalence Principle with Free-Fall
Rotating Gyroscopes}
\author{J. LUO$^{\text{a *}}$, Y. X. Nie$^{\text{b}}$, Y. Z. Zhang$^{\text{c,d}}$,
Z. B. Zhou$^{\text{a}}$}
\address{$^{\text{a}}$Department of Physics, Huazhong Univeristy of Science and\\
Technology, Wuhan 430074, China\\
$^{\text{b}}$Institute of Physics, Chinese Academy of Sciences, Beijing\\
100080, China\\
$^{\text{c}}$CCAST(World Lab.), P. O. Box 8730, Beijing 100080\\
$^{\text{d}}$Institute of Theoretical Physics, Chinese Academy of Sciences,\\
Beijing 100080, China**}
\date{June 20, 2001}
\maketitle

\begin{abstract}
The differential acceleration between a rotating mechanical gyroscope and a
non-rotating one is directly measured by using a double free-fall
interferometer, and no apparent differential acceleration has been observed
at the relative level of 2$\times $10$^{\text{-6}}$. It means that the
equivalence principle is still valid for rotating extended bodies, i.e., the
spin-gravity interaction between the extended bodies has not been observed
at this level. Also, to the limit of our experimental sensitivity, there is
no observed asymmetrical effect or anti-gravity of the rotating gyroscopes
as reported by hayasaka et al.
\end{abstract}


\section{Introduction}

It is well known that spin-interactions of elementary particles, spin-orbit
coupling and spin-spin coupling, have been studied in both theory and
experiment for long time. Furthermore, gravitational couplings (i.e. the
spin-gravitoelectric coupling \cite{R1,R2} and the spin-gravitomagnetic
coupling \cite{R3,R4}) and spin-rotation coupling \cite{R5,R6,R7} between
intrinsic spins and rotating bodies have been also investigated for long
time (see, e.g., \cite{R8}).

However, the status of research for rotation (spin)-coupling
between macroscopic rotating bodies is greatly different. The
spin-orbit coupling for motion of mechanical gyroscope has been
already well known in Newton's mechanics. With the exception of
the spin-orbit coupling, on the other hand, Einstein theory of
general relativity also predicts the spin-gravitational coupling
of mechanical gyroscope, which has been investigated by many
authors, e.g. see Ref. 8. In particular, the Stanford Gravity
Probe B (GPB)group has theoretically studied for long time on
these types of gravitomagnetic effects and planed to perform a
satellite orbital experiment in order to seek the couplings of
rotor spin to Earth spin and rotor spin to rotor orbit \cite{R9}.
As pointed out by Zhang et al. \cite{R10}, however, the mechanical
gyroscope spin is essentially different from the intrinsic spin of
elementary particle. In fact, an extended body could have two
different types of motion, i.e. orbit motion (the motion of the
center-of-mass) and rotation. Thus a extra force (or force
moment), which could come from the spin-spin (i.e.
rotation-rotation) coupling between rotating macroscopic bodies,
might change the three types of motion for the rotating bodies:
(i) spin precession (i.e. a change of spin direction), (ii) a
change of the rotation rate, and (iii) a change of the motion of
the center-of-mass. It is known that general relativity (GR) only
predicts (i), i.e. spin precession. While any possible connections
of GR with (ii) and (iii) are now still open problems. Thus the
Stanford GPB project simply includes a measurement of the spin
precession rather than the (ii) and (iii). In addition, although
other gravitational theories, such as the gauge theories of
gravitation with torsion \cite{R11}, seem to include spin-spin
coupling of fluid, it is difficult to discuss the spin-interaction
between rotating rigid balls within the framework of these
theories. For this reason, Zhang et al. recently developed a
phenomenological model for the rotation-rotation interaction
between the rotating rigid balls \cite{R10}, which predicts (iii),
i.e. the effect of the coupling, gyroscope spin to Earth spin, on
the orbital acceleration of the gyroscope free-falling in Earth' s
gravitational field. In this sense this type of spin-spin coupling
would violate the equivalence principle (EP) for the free-fall
gyroscopes.

EP, as one of the fundamental hypotheses of Einstein's general relativity,
has been tested by many experiments \cite{R12,R13,R14,R15,R16,R17,R18}.
Recently, some different tests of EP for gravitational self-energy \cite{R19}
and spin-polarized macroscopic objects \cite{R20,R21} have been reported.
However, in all of the experiments as well as the Satellite Test of the
Equivalence Principle (STEP) and the Galileo Galilei (GG) space projects as
well as the MICROSCOPE space mission \cite{R22,R23,R24}, it is non-rotating
bodies that are used. In addition, as pointed out above, although a
gyroscope is used in the Stanford Gravity Probe B project, only the
precession of the gyroscopic spin is to be observed, which is irrelevant to
the orbital motion.

Some relevant experiments have been performed by use of mechanical
gyroscopes and give contradictory results \cite{R25,R26,R27,R28,R29,R30}. In
particular, the observations in these experiments were made by means of beam
balance, and so only the gravity and its reacting force were working, which
is irrelevant to inertial force. Therefore, this type of experiment is
simply a test of statics independent of EP.

Recently, Hayasaka et al. investigated the effect of a rotating gyroscope on
the fall-acceleration by comparing the fall-times of the gyroscope with
differential rotating sense using the time-counter combined with three
couples of the laser-emitters and receivers \cite{R31}. Their experimental
data show that the gravity acceleration of the right-rotating rotor at 18000
rpm is smaller than that of non-rotating one at the relative level of 10$^{%
\text{-4}}$ , and the gravity acceleration of the left-rotating rotor almost
identical with that of the non-rotating (i.e. an asymmetric coupling). But
the phenomenological theory for rotating rigid balls in Ref. 10 predicts a
symmetric spin-spin coupling which is in the order of 10$^{\text{-14}}$ much
less than the observation in Ref. 31. As pointed out above, this type of
free-fall experiment is a test of dynamics, which is closely related to EP.
And hence, it is necessary to do a new dynamic test of EP by use of
free-fall gyroscopes.

In this article, we shall report a new dynamic test of the spin-spin
coupling between a gyroscope and the Earth. Based on the theoretical model
in Ref. 10, a dimensionless parameter representing the strength of violation
of EP can be defined as follows:

\begin{equation}
\eta _{\text{s}}=\frac{\Delta g}{g}=\kappa \left( \frac{\stackrel{%
\rightharpoonup }{S}_{\text{1}}\cdot \stackrel{\rightharpoonup }{S}_{\text{e}%
}}{Gm_{\text{1}}M_{\text{e}}R_{\text{1}}}-\frac{\stackrel{\rightharpoonup }{S%
}_{\text{2}}\cdot \stackrel{\rightharpoonup }{S}_{\text{e}}}{Gm_{\text{2}}M_{%
\text{e}}R_{\text{2}}}\right) \text{ ,}  \label{Eq1}
\end{equation}
where {\it G} is the Newtonian gravitational constant, {\it m}$_{\text{1}}$,
{\it m}$_{\text{2}}$ and {\it M}$_{\text{e}}$ are the masses of the two
gyroscopes and the Earth, respectively, and $\stackrel{\rightharpoonup }{S}_{%
\text{1}}$, $\stackrel{\rightharpoonup }{S}_{\text{2}}$, and $\stackrel{%
\rightharpoonup }{S}_{\text{e}}$ are the spin angular momentums of them
correspondingly, {\it R}$_{\text{1}}$ and {\it R}$_{\text{2}}$ are the
distances between the centers of the two gyroscopes and the Earth,
respectively, and the parameter $\kappa $ represents the universal coupling
factor for the spin-spin interaction. Therefore, in a double free-fall (DFF)
experiment, in which two gyroscopes with differential rotating senses drop
freely, an observed non-zero value of $\eta _{\text{s}}$ would imply
violation of the EP or existence of spin-spin force between the gyroscope
and the Earth.

\section{Experimental Description}

The schematic diagram of the DFF experiment is shown in Fig. 1. A
frequency-stabilized He-Ne laser beam (633 nm) with the relative length
standard of 1.3$\times $10$^{\text{-8}}$ is split by two beam splitters and
sent vertically to the two corner-cube-retroreflectors (CCRs) fixed on the
bottoms of the test masses, respectively, and then combined again and forms
interference fringes on a 12 ns-response-time photodiode ({\it RS Ltd.,
OSD15-5T}). The differential vertical displacement of both test masses, the
gyroscopes with differential rotating senses, is continuously monitored by
the interferometer and sampled by means of a 10 MHz data-acquisition-card (%
{\it Gage Ltd., Cs1250}) combined with an external rubidium atomic clock (%
{\it SRS Inc., SR620}), which provides a relative time standard of 10$^{%
\text{-10}}$, and then stored in a computer. The test masses are freely
dropped in two 12 m-high vacuum tubes of about 20 $\sim $ 50 mPa. Compared
with the SFF experiment employed by Hayasaka et al., the DFF scheme can
minimize the environmental noises such as the tides, gravity gradient,
seismic noise, and air damping and so on, because the differential mode
design can suppress some common errors of both falling objects.

As we known, the sensitivity of such a Galilean experiment in which both
dropping objects are put side-by-side is limited by the alignment of the
beam propagation away from the vertical line \cite{R17}. For example, an
error in the verticality of 5$^{\prime \prime }$ will contribute an
uncertain differential acceleration of 0.3 $\mu $Gal (1 Gal = 1 cm/s$^{\text{%
2}}$). A proposed method to reduce this error is to locate the dropping
masses directly one above the other, but the design and operation would be
very complicate. However, in order to test the asymmetrical gravity
acceleration effect of 10$^{\text{-4}}$ as reported by Hayasaka et al, the
side-by-side setup is employed here, and the two test masses are separated
horizontally (south-north) by 480 mm. This design is very convenient for us
to drive the gyroscopes and release them.

Each of the two test masses consists of a steel gyroscope with a mass of
420.0$\pm $2.5 g, a diameter of about 55 mm and a height of about 32 mm, a
CCR of 76.4$\pm $0.4 g and a diameter of 40 mm as well as an outer aluminum
frame of 159.4$\pm $0.9 g. Tinned copper wires with a diameter of 0.25 mm
are used to suspend the test masses and melt by an instantaneous large
current (
\mbox{$>$}%
150 A) provided by a capacitor array, and then the test masses are released
and drop freely \cite{R32}. A DC three-phase motor is used to drive one of
the gyroscopes, and the other is in non-rotating status. The rotating speed
of the gyroscope can be adjusted by changing the input voltage of the motor.
Simultaneous measurement of the driving frequency of the motor and the
rotating rate of the gyroscope rotor in a vacuum container of about 3 Pa
showed that the rotating frequency of the rotor is equal to that of the
motor with an uncertainty of 1\%. It is useful for recording the rotating
speed of the gyroscope without adding an external measurement system in the
vacuum chamber. The rotating speed of the gyroscope is kept at (17000$\pm $%
200) rpm. A mechanical claw is used to grasp the test mass during the
speedup progress of the gyroscope, and it is then loosed when the gyroscope
runs normally. The free-fall test masses are captured by two 1.2 m-length
tubes with an assembly of thin rubber-rings and aluminum foils,
respectively. Because of the lack of a return mechanism, which could reset
the dropping objects under the vacuum condition, we have to open the vacuum
tubes after each free-fall measurement.

The diameter of the laser beam is kept in a range of 3.0 $\sim $ 3.2 mm by a
two-lens collimation assembly during 20 m optical length so that the beam
wavefront effect can be neglected here. The differential radiation pressure
on the test masses is less than 3$\times $10$^{\text{-4}}$ ${\sc \mu }$Gal
for 0.5 mW laser power used here. The angles of the beam aligned with the
local vertical are monitored by a telescope combined with two horizontal oil
references, and then fed back to align the beam splitters by four fine
screws. The aligned verticality is kept within 50$^{\prime \prime }$\ for
each laser beam, the maximum uncertainty of the differential acceleration
due to the aligned verticality is 30 ${\sc \mu }$Gal.

The test mass with a non-rotation rotor is released about 3 ms before the
other with a left- or right-rotating rotor in order to obtain an
interference fringe rate of about 100 kHz by mean of two differential relay
switches. The amplitude spectrum of the seismic noise in our laboratory is
about $10^{-9}/(f/$Hz$)^{2}$ m{\tt /}$\sqrt{\text{Hz}}$\ \cite{R33}, which
will contribute an uncertainty of about 1 ${\sc \mu }$Gal to the final
experiment result.

The sample data in each free-fall are processed as following steps. First,
the DC-offset and the amplitude of each interference fringe are determined
from the original time-voltage data \{$t_{\text{i}},V_{\text{i}}$\}. Second,
by calculating an inverse function of the fringe using the DC-offset and the
amplitude determined, we can transform the data \{$t_{\text{i}},V_{\text{i}}$%
\} into the time-differential displacement data \{$t_{\text{i}},\Delta z_{%
\text{i}}$\}. Finally, the data \{$t_{\text{i}},\Delta z_{\text{i}}$\} are
fitted by a parabolic trajectory perturbed with a linear vertical gravity
gradient ${\it \gamma }$. The differential displacement between both test
masses is given by the equation as follows

\begin{equation}
\Delta z=\Delta z_{0}+\Delta v_{0}t+(\Delta g+{\it \gamma }\Delta
z_{0})t^{2}/2+\Delta v_{0}{\it \gamma }t^{3}/6\text{ ,}  \label{Eq2}
\end{equation}
where unknown parameters $\Delta z_{0}$, $\Delta v_{0}=g\Delta t_{0}$, and $%
\Delta g$ are the initial differential vertical displacement, velocity, and
acceleration at the same height, respectively. It is evident that the
initial differential displacement, which includes their original suspending
difference and descent height due to the release time-delay $\Delta t_{0}$,
has to be measured accurately. Here the suspending height difference of both
test masses is less than 1 mm, and their descent height due to 3 ms delay is
about 50 ${\sc \mu }$m. In this case, the vertical gravity gradient effect
is about 0.3 ${\sc \mu }$Gal. In addition, it is noted that the fitting
initial time difference, which is here defined as the time difference of the
fitting initial data away from the real release time of the latter test
mass, will contribute an uncertain acceleration difference due to the
coupling between the initial differential velocity and the vertical gravity
gradient. In general, the fitting initial time difference should be kept
below 0.1 s for 1 $\mu $Gal uncertainty.

A known systematic error due to the finite speed of light is given by \cite
{R34}

\begin{equation}
\Delta g/g\text{ }\simeq \text{ }3\Delta v_{0}/C\text{ ,}  \label{Eq3}
\end{equation}
and the correction is about 0.3 ${\sc \mu }$Gal in our experiment. Another
systematic error due to residual gas drag could be calculated as follows
\cite{R35}

\begin{equation}
\Delta g/g=A\Delta v_{0}p\sqrt{8{\it \mu }/(\pi RT)}/(4mg)\text{ ,}
\label{Eq4}
\end{equation}

where $A$ ( $\approx 170$ cm$^{\text{2}}$) is the total surface area of the
test mass, ${\it \mu }$\ and $R$ are the molecular weight of residual gas
and the gas constant, $m$ is the mass of the falling object, $T$ is the
temperature, and $p$ is the residual pressure. The uncertain acceleration
due to the drag effect is less than 5 ${\sc \mu }$Gal at $p$ = 50 mPa and $T$
= 300 K.

Variation of the magnetic flux density is within 0.1 Gauss near the right-,
left-, or non-rotating rotor, and the geomagnetic flux density is about 0.4
Gauss here. The estimation shows that the effect of the geomagnetic field on
the steel rotor is at the level of 10$^{\text{-10}}$ Gal.

An acceleration difference due to interaction between a possible horizontal
velocity difference $\Delta v_{\text{h}}$ and rotation of the Earth is given
by

\begin{equation}
\Delta g/g=2\Delta \stackrel{\rightharpoonup }{v}_{\text{h}}\times \stackrel{%
\rightharpoonup }{\it \Omega }\leq 2\Delta v_{\text{h}}{\it \Omega }\cos
\lambda \text{ ,}  \label{Eq5}
\end{equation}
where ${\it \Omega }$ is the angular frequency of the Earth's rotating, and $%
{\it \lambda }$ ( $\simeq $ 30 degree) is the latitude of our laboratory.
The $\Delta v_{\text{h}}$ is estimated smaller than 4.3 mm/s according to
interference intensity of the two interference beams reflected from the CCRs
versus the falling length (6 mm deviation for 10 m-fall height). Therefore,
the uncertain acceleration due to the procession effect is less than 54 $\mu
$Gal. It means that the horizontal velocity difference would have to be
monitored in the further experiment with a higher precision.

A possible lifting force for a rotating rotor due to the residual gas flow's
circulation can be calculated based on the Zhukovskii's theorem as follows
\cite{R36}

\begin{equation}
\stackrel{\rightharpoonup }{F}_{\text{lift}}=m\stackrel{\rightharpoonup }{a}%
_{\text{lift}}=-2\rho _{\text{gas}}\stackrel{\rightharpoonup }{V}\times
\stackrel{\rightharpoonup }{\omega }\text{ ,}  \label{Eq6}
\end{equation}

where $\stackrel{\rightharpoonup }{V}$ is the velocity of the rotating
rotor, $\stackrel{\rightharpoonup }{\omega }$ ( $\sim $~17000 rpm) is the
angular velocity of the rotating rotor, and $\rho _{\text{gas}}$ is the
residual gas density in the vacuum tube. Because the interferometric
measurement here is nearly insensitive to the horizontal motions of the two
test masses, the lifting effect on the vertical acceleration difference
would be zero if $\stackrel{\rightharpoonup }{\omega }$ was exactly along
the vertical axis. The maximum uncertain rotation direction of the rotating
rotor away the vertical axis is estimated within 2.4 mrad, thus a possible
vertical acceleration difference between the rotors due to the gas flow's
lifting is at the level of 10$^{\text{-10}}$ Gal, which can be neglected
here.

\section{Experimental Results}

A typical voltage output from the photodiode is shown in Fig. 2. Figure 2(a)
is the intensity curve of the interference fringe as the first dropping
object (non-rotating here) is released, and the rate of the fringe increases
with the falling of the non-rotating test mass until the other is also
released. As both test masses drop freely, the rate of the fringe is
modulated by their acceleration difference or the noises, as shown in Fig.
2(b).

Figure 3 lists 3 sets of 5 measurements each of N-L, N-R, and N-N, where L,
R, and N represent left-, right-, and non-rotating, respectively. The
uncertain differential acceleration of each free-fall comes to the level of
1000 ${\sc \mu }$Gal, while the fitting standard deviation ($\pm 1{\it %
\sigma }$) is only a few ${\sc \mu }$Gal, but it is noted that the
uncertainty is independent of their rotating senses. Statistical result
shows that relative uncertainty of the differential acceleration between the
non- and left-rotating test masses is $\Delta g_{\text{N-L}}/g=(0.90\pm
0.94)\times 10^{-6}$, and that between the non- and right-rotating is $%
\Delta g_{\text{N-R}}/g=(0.67\pm 1.92)\times 10^{-6}$. They are almost the
same as the background limit of $\Delta g_{\text{N-N}}/g=(0.56\pm
1.44)\times 10^{-6}$.

Summarizing the data obtained in Ref. 25, the weight loss, resulting from
the mass reduction or the acceleration decrease, for right-rotating around
the vertical axis is approximately formulated by Hayasaka and Takeuchi, in
units of dynes, as follows

\begin{equation}
\Delta W({\it \omega })={\it \alpha }\text{{\it m}}r_{\text{eq}}{\it \omega }%
\text{ ,}  \label{Eq7}
\end{equation}
where $m$ is the mass of rotor (in g), ${\it \omega }$\ is the angular
frequency of rotation (in rad/s), and $r_{\text{eq}}$ is the equivalence
radius (in cm), defined as follows

\begin{equation}
mr_{\text{eq}}=\int \int {\it \rho }(r,z)2\pi r^{2}drdz\text{ ,}  \label{Eq8}
\end{equation}
where ${\it \rho }(r,z)$ is the density of the rotor materials. Their
experimental result shows that the factor ${\it \alpha }$ is about $2\times
10^{-5}/${\sc s}. Considering the generalization of the possible anomalous
weight change of the rotating gyroscopes, the possible weight loss of the
two rotating directions of the gyroscope could be given as follows \cite
{R29,R37}

\begin{equation}
\Delta W({\it \omega })={\it \beta }I{\it \omega }\text{ ,}  \label{Eq9}
\end{equation}
where $I$ is the inertia moment of the rotating rotor, ${\it \beta }$ could
be considered as a factor dependent upon the anomalous effect. Based on the
above formulas, all reported experimental tests of the anomalous effect are
tabulated in Table I as suggested by Newman \cite{R38}. It is noted that
some unknown parameters are calculated according to a uniform composition
rotor assumption.

>From the results of our DFF experiment, there is no apparent differential
acceleration between the rotating and non-rotating test masses within our
experimental limits. Therefore, we can conclude that the differential
acceleration between the rotating and non-rotating gyroscopes is almost 2
orders of magnitude smaller than reported in Ref. 31, and the differential
acceleration effect between the right- and left- versus the non-rotating has
not observed in our experiment at the relative level of 2$\times $10$^{\text{%
-6}}$. It means that EP is still valid for extended rotating bodies, and the
spin-spin interaction between the rotating extended bodies has not been
observed at this level. And then, according to the Eq. (1) and the
approximately uniform sphere mode of the Earth, it can be concluded that $%
\kappa \leq $2$\times $10$^{\text{-18 }}$kg$^{\text{-1}}$, which sets an
upper limit for the spin-spin interaction between a rotating extended body
and the Earth.

\section{Discussion}

A large limitation in our experiment comes from the friction coupling
between the rotating rotor and the frame of the test mass. The friction
coupling not only causes a high-frequency mechanical vibration of the CCR at
the frequency of the rotating rotor, but also results in a slowly rotating
motion of the frame, which frequency is about 1 Hz. Another main limitation
had been proved to come from the outgassing effect of the vacuum pump with a
full rated pumping speed 1500 L/s due to the asymmetrical outgassing for the
two tubes here. It is hoped that the sensitivity of our DFF experiment could
be improved by one or two orders in the near future, and the upper limit of
the dependent factors $\alpha $ or $\beta $ could be improved to 10$^{\text{%
-9}}$. Therefore, the new EP for the rotating extended bodies could be
tested at the same level correspondingly.

{\bf Acknowledgments} We are grateful to Prof. W. R. Hu and Prof. R. D.
Newman for their discussion and useful suggestion. This work was supported
by the Ministry of Science and Technology of China under Grant No: 95-Yu-34
and the National Natural Science Foundation of China under Grant No:
19835040.

* Email address: junluo@public.wh.hb.cn.

** Mailing address of Y. Z. Zhang (e-mail: yzhang@itp.ac.cn)

\begin{figure}[tbp]
\caption{Schematical diagram of a free-fall interferomater used to measure
the differential acceleration between two gyroscopes with differential
rotating senses.}
\label{Figure 1}
\end{figure}

\begin{figure}[tbp]
\caption{Interference fringe intensity as the first test mass is released.
The fringe rate increases with its falling; (b) Interference fringe
intensity as both the test masses drop freely.}
\label{Figure 2}
\end{figure}

\begin{figure}[tbp]
\caption{Statistical result of the relative differential acceleration
between two test masses with different rotating senses. L, R, and N
represent left, right-, and non-ratating, respectively, L$_{\text{sta}}$, R$%
_{\text{sta}}$, and N$_{\text{sta}}$ represent the statistical values of the
corresponding differential acceleration, and the error bars denote $\pm
1\sigma .$}
\label{Figure 3}
\end{figure}

\begin{figure}[tbp]
\caption{Summary of test experiments of anomalous weight change of the
rotating rotors.}
\label{Table 1}
\end{figure}

\begin{tabular}{cccccccccc}
\hline\hline
Experiment & Method & $M$ & $D$ & $r_{\text{eq}}$ & $I$ & ${\it \omega }_{%
\text{max}}$ & $\Delta W$ & ${\it \alpha }$ & ${\it \beta }$ \\
&  & (g) & (cm) & (cm) & (g$\cdot $cm$^{\text{2}}$) & (rpm) & (dyn) & (s$^{%
\text{-1}}$) & (cm$^{\text{-1}}$s$^{\text{-1}}$) \\ \hline
Hayasaka \& Takeuchi & BB & 140 & 5.2 & 1.85$^{\text{a}}$ & 473$^{\text{b}}$
& 13000 & 7.6 & 2.14$\times $10$^{\text{-5}}$ & 1.17$\times $10$^{\text{-5}}$
\\
&  & 175 & 5.8 & 2.26$^{\text{a}}$ & 736$^{\text{b}}$ &  & 11.7 & 2.17$%
\times $10$^{\text{-5}}$ & 1.16$\times $10$^{\text{-5}}$ \\
Faller et al. & BB & 451 & 5.1 & 1.70$^{\text{b}}$ & 1466$^{\text{b}}$ & 6000
&
\mbox{$<$}%
0.39 & $\leq $ 8.14$\times $10$^{\text{-7}}$ & $\leq $ 4.25$\times $10$^{%
\text{-7}}$ \\
Quinn \& Picard & BB & 330 & 4.0 & 1.33$^{\text{b}}$ & 660$^{\text{b}}$ &
8000 &
\mbox{$<$}%
0.06 & $\leq $ 1.60$\times $10$^{\text{-7}}$ & $\leq $ 1.06$\times $10$^{%
\text{-7}}$ \\
Nitschke \&Wilmarth & BB & 142 & 3.84 & 1.28$^{\text{b}}$ & 328$^{\text{a}}$
& 22000 &
\mbox{$<$}%
0.07 & $\leq $ 1.64$\times $10$^{\text{-7}}$ & $\leq $ 0.91$\times $10$^{%
\text{-7}}$ \\
Imanishi et al. & BB & 129 & 5.0 & 1.94$^{\text{a}}$ & 551$^{\text{a}}$ &
11000 &
\mbox{$<$}%
0.32 & $\leq $ 1.12$\times $10$^{\text{-6}}$ & $\leq $ 5.10$\times $10$^{%
\text{-7}}$ \\
Hayasaka et al. & SFF & 175 & 5.8 & 1.93$^{\text{b}}$ & 970$^{\text{a}}$ &
18000 & 24.9 & 3.90$\times $10$^{\text{-5}}$ & 1.36$\times $10$^{\text{-5}}$
\\
Luo et al. & DFF & 420 & 5.49 & 1.83$^{\text{b}}$ & 1582$^{\text{b}}$ & 17000
&
\mbox{$<$}%
0.80 & $\leq $ 5.89$\times $10$^{\text{-7}}$ & $\leq $ 2.86$\times $10$^{%
\text{-7}}$ \\ \hline\hline
\end{tabular}

$^{\text{a}}$ data provided by the corresponding reference.

$^{\text{b}}$ data calculated according to the assumption of a unifrom
composition rotor.

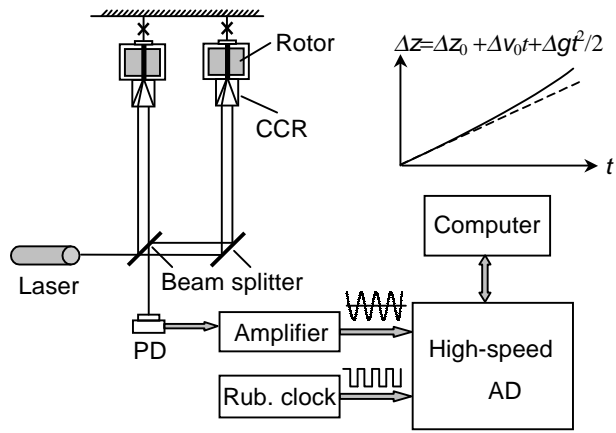
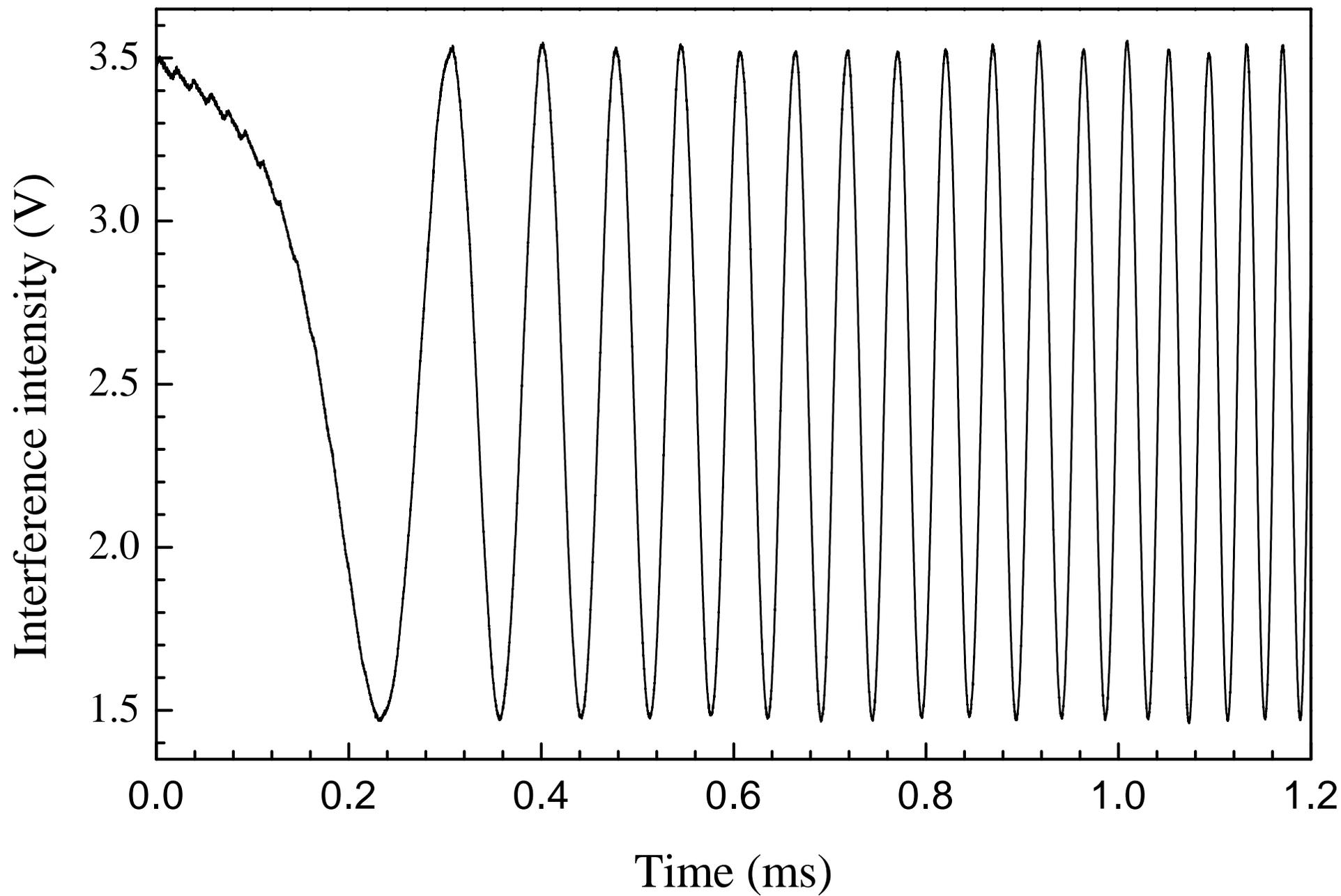
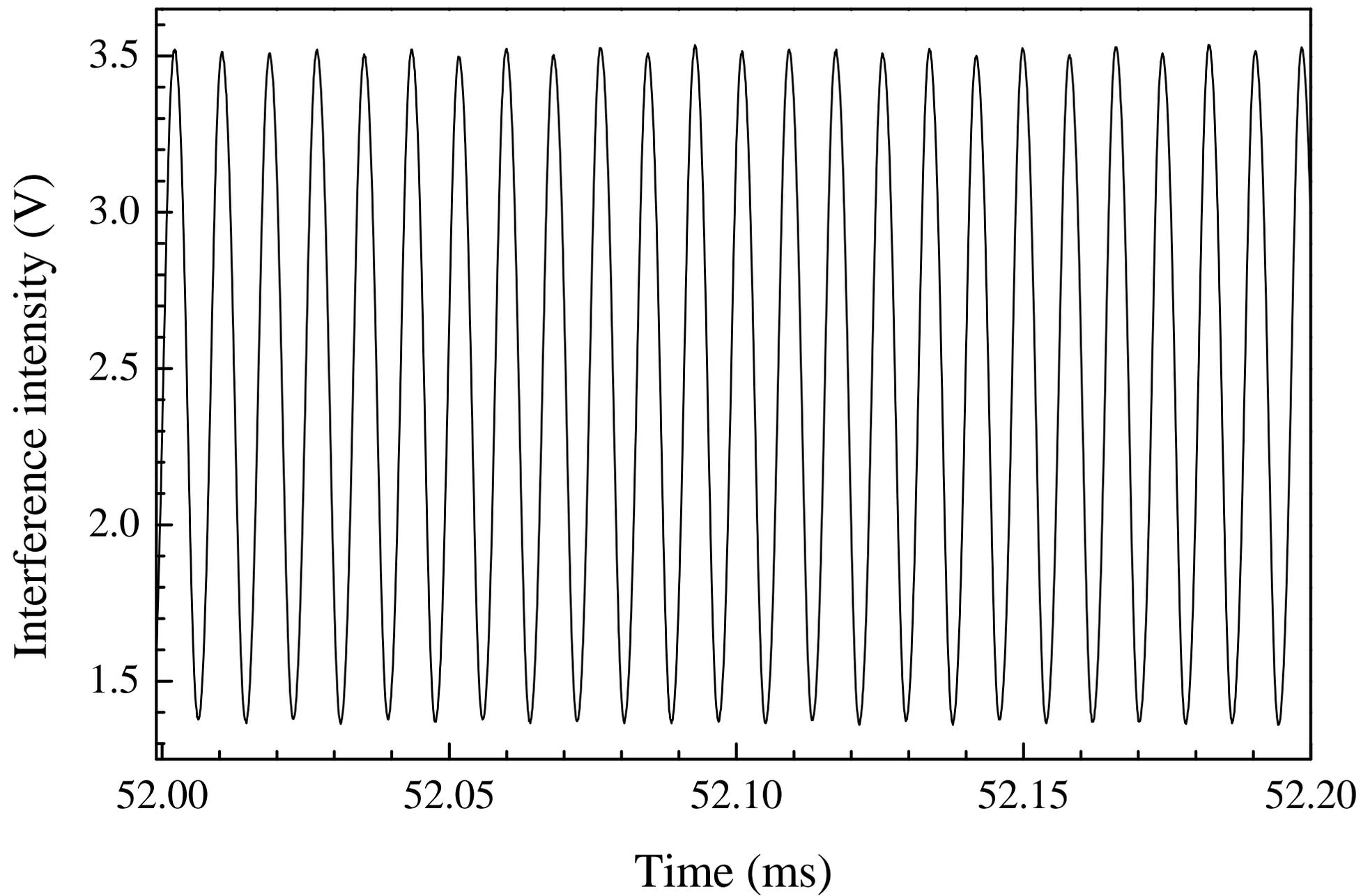
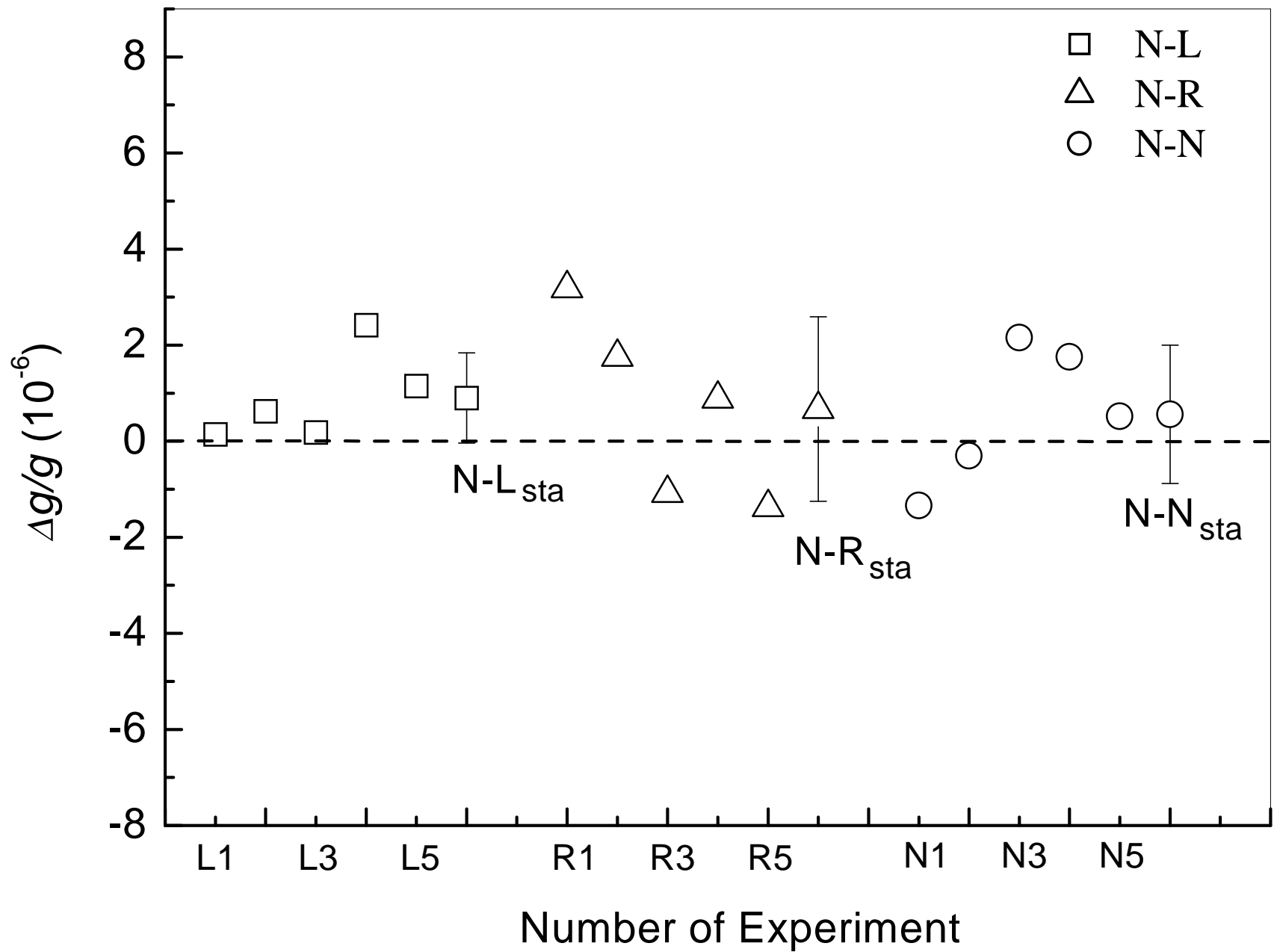
\end{document}